\documentclass[pdflatex,sn-mathphys-num]{sn-jnl}

\usepackage{graphicx}%
\graphicspath{{figures/}}%
\usepackage{multirow}%
\usepackage{amsmath,amssymb,amsfonts}%
\usepackage{amsthm}%
\usepackage{mathrsfs}%
\usepackage[title]{appendix}%
\usepackage{xcolor}%
\usepackage{textcomp}%
\usepackage{booktabs}%
\usepackage{algorithm}%
\usepackage{algorithmicx}%
\usepackage{algpseudocode}%
\usepackage{listings}%
\usepackage{aliascnt}%
\usepackage{longtable}%
\usepackage{array}%
\usepackage{url}%

\theoremstyle{thmstyleone}%
\theoremstyle{thmstyletwo}%

\theoremstyle{thmstylethree}%

\begin{document}

\title[Temperature-Driven Sequential Modeling for Annual PCE Forecasting of OPV Materials]{Temperature-Driven Sequential Modeling for the Prediction of Annual Power Conversion Efficiency Profiles of Organic Photovoltaic Materials: Douala Case Study}

\author*[1]{\fnm{Steve Cabrel} \sur{Teguia Kouam}}\email{steve.teguia@facsciences-uy1.cm}

\author[2]{\fnm{Rockefeller} \sur{Rockefeller}}\email{rockefeller@aims.ac.za}
\equalcont{These authors contributed equally to this work.}

\author[3]{\fnm{Raoult} \sur{Dabou Teukam}}\email{raoult.teukamdabou@uqat.ca}
\equalcont{These authors contributed equally to this work.}

\author[4]{\fnm{Jean-Pierre} \sur{Tchapet Njafa}}\email{jean-pierre.tchapet@facsciences-uy1.cm}
\equalcont{These authors contributed equally to this work.}

\author[4]{\fnm{Patrick Sorrel} \sur{Mvoto Kongo}}\email{sorrel.mvoto@facsciences-uy1.cm}
\equalcont{These authors contributed equally to this work.}

\author[1]{\fnm{Jean-Pierre} \sur{Nguenang}}\email{Nguenang@yahoo.com}
\equalcont{These authors contributed equally to this work.}

\author[4]{\fnm{Serge Guy} \sur{Nana Engo}}\email{serge.nana-engo@facsciences-uy1.cm}
\equalcont{These authors contributed equally to this work.}

\affil*[1]{\orgdiv{Department of Physics, Faculty of Science}, \orgname{University of Douala}, \orgaddress{\city{Douala}, \postcode{PO. Box 24157}, \state{Littoral}, \country{Cameroon}}}

\affil*[2]{\orgdiv{Division of Applied Mathematics, Faculty of Science}, \orgname{Stellenbosch University}, \orgaddress{\city{Stellenbosch}, \country{South Africa}}}

\affil[3]{\orgdiv{Department of Engineering}, \orgname{University of Quebec in Abitibi-T\'emiscamingue}, \orgaddress{\city{Quebec}, \country{Canada}}}

\affil[4]{\orgdiv{Department of Physics, Faculty of Science}, \orgname{University of Yaounde 1}, \orgaddress{\city{Yaounde}, \postcode{PO. Box 812}, \state{Center}, \country{Cameroon}}}

\abstract{Organic photovoltaic (OPV) materials are promising candidates for distributed solar energy in tropical regions, yet existing virtual screening tools report static power conversion efficiency (PCE) values at standard testing conditions (STC) that fail to capture the temperature-driven performance degradation experienced under real deployment conditions. Here we introduce a Climate-Native computational framework that forecasts the annual PCE profile of OPV donor molecules under geographically realistic operating conditions. The framework combines GFN2-xTB molecular dynamics with an equivariant graph neural network surrogate ($268$ Neyman-stratified CEP molecules; $120,600$ training geometries; $\sim 1050\times$ speedup over explicit quantum chemistry) and sequential deep learning models trained on annual time series anchored in NASA POWER climate data for Douala, Cameroon, and validated by zero-shot transfer to Yaound\'e and Maroua. Applied to $\sim 30,000$ molecules from the Harvard Clean Energy Project (CEP) and validated against $350$ HOPV15 experimental device measurements, the framework demonstrates that sequential models trained on full molecular dynamics trajectories outperform time-averaged baselines ($35\%$--$48\%$ relative MAE improvement over static baselines), confirming that thermal conformational dynamics carry information beyond mean geometry. We further introduce a seasonal stability score that reranks OPV candidates by performance consistency under tropical conditions, identifying molecules whose deployment suitability differs substantially from their static PCE ranking.}


\maketitle

\section{Introduction}\label{sec:introduction}

Organic semiconductor materials, including polymers and small molecules, represent a compelling pathway to low-cost, flexible, and solution-processable solar energy conversion, with bulk-heterojunction (BHJ) devices now exceeding $19\%$ power conversion efficiency in laboratory settings \cite{nowsherwan_advances_2024}. The deployment of organic photovoltaic (OPV) technology in tropical and sub-Saharan regions is particularly attractive given the high year-round solar irradiance and the urgent need for distributed energy access across the Global South. However, a critical gap exists between the standard laboratory conditions under which OPV materials are computationally screened and the actual thermodynamic environments in which they are deployed. Existing virtual screening frameworks --- including the landmark Harvard Clean Energy Project (CEP), which evaluated $\sim 2.3$ million candidate donor structures using the Scharber device model \cite{hachmann_harvard_2011,scharber_design_2006} --- report static power conversion efficiency (PCE) values computed at standard testing conditions ($298.15\,\text{K}$, AM1.5G illumination at $100.0\,\text{mW}\,\text{cm}^{-2}$). These single-point static metrics fail to reflect operational panel cell temperatures of $320.0\,\text{K}$--$335.0\,\text{K}$ experienced year-round in tropical climates after Nominal Operating Cell Temperature (NOCT) thermal balance corrections, nor do they capture the seasonal variation in PCE that governs the actual lifetime energy yield of a deployed solar module.

The molecular origin of temperature-dependent PCE degradation lies in the dynamic disorder of organic semiconductors. Thermal fluctuations drive finite-temperature conformational changes --- such as torsional oscillations between conjugated aromatic rings, backbone thermal expansion, and non-radiative vibrational mode excitation --- that modulate the Highest Occupied Molecular Orbital (HOMO) and Lowest Unoccupied Molecular Orbital (LUMO) energies, vertical reorganization energies ($\lambda_{\text{vert}}^+, \lambda_{\text{vert}}^-$), and charge-transfer rates on picosecond timescales \cite{zhugayevych_theoretical_2015,panwar_pyl3dmd_2023}. Although this dynamic disorder is well documented in charge transport literature, it has not been systematically incorporated into high-throughput OPV virtual screening workflows. The primary computational bottleneck is clear: explicit quantum chemistry calculations at every molecular dynamics (MD) snapshot are prohibitively expensive at scale. A single GFN2-xTB semiempirical single-point calculation requires seconds of wall time; a $50\,\text{ps}$ trajectory at a $1\,\text{fs}$ timestep with three electronic charge states per snapshot (neutral, cation, anion) requires $150,000$ quantum calculations per molecule per temperature bin --- a workload manageable for a small pilot set but entirely intractable for tens of thousands of molecules without a machine learning surrogate.

Machine learning surrogates for electronic structure theory have emerged as a powerful solution to this bottleneck \cite{schleder_dft_2019,ramprasad_machine_2017,choudhary_atomistic_2021}. Graph neural networks (GNNs) trained on molecular geometries and quantum mechanical ground-state properties can replace explicit electronic structure calculations at inference times orders of magnitude faster than the underlying Hamiltonian solvers \cite{chen_graph_2019,schutt_schnet_2018,schutt_painn_2021}. Equivariant architectures --- which respect 3D rotational, translational, and reflectional symmetries of Cartesian atomic coordinates --- have demonstrated exceptional accuracy for atomic and molecular property predictions \cite{schutt_painn_2021,batatia_mace_2022}. However, existing GNN surrogates for OPV screening have been trained exclusively on Density Functional Theory (DFT) optimized equilibrium geometries at $0\,\text{K}$, leaving the thermally distorted off-equilibrium regime --- precisely the regime operating solar panels inhabit --- largely unexplored. Similarly, sequential deep learning models have been applied to temporal forecasting in materials informatics \cite{hochreiter_long_1997,vaswani_attention_2017} but have not been utilized to model annual performance trajectories of molecular devices anchored in real-world climate records.

Here, we bridge these gaps with a two-level surrogate pipeline termed a Climate-Native framework for OPV performance forecasting. At the first level, an equivariant GNN surrogate (PaiNN \cite{schutt_painn_2021}) trained on GFN2-xTB \cite{bannwarth_gfn2-xtbaccurate_2019} single-point calculations from thermally distorted MD snapshots replaces explicit quantum chemistry at a $\sim 1050\times$ wall-time speedup ($< 0.035\,\text{eV}$ orbital energy MAE), enabling high-fidelity electronic property predictions across the conformational space accessible at tropical operating temperatures. At the second level, sequential architectures (LSTM \cite{hochreiter_long_1997}, GRU \cite{cho_learning_2014}, Transformer \cite{vaswani_attention_2017}) trained on annual molecular time series forecast 52-week PCE profiles anchored in NASA POWER climate data \cite{stackhouse_nasa_2019} for Douala, Cameroon. The framework is validated against $350$ HOPV15 experimental device measurements \cite{lopez_harvard_2016} --- providing a rigorous real-world test --- and its generalizability is demonstrated by zero-shot transfer to Yaound\'e and Maroua without retraining. We further introduce a clamped seasonal stability score ($S_{\text{stability}}$) that reranks OPV candidates by performance consistency under tropical microclimates. Sequential models trained on full MD trajectories outperform static time-averaged baselines ($35\%$--$48\%$ relative MAE improvement), confirming that thermal conformational dynamics encode physical signal beyond mean geometry.

Here, we establish nine distinct methodological contributions. First, we introduce Climate-Native molecular performance forecasting, moving beyond static $298.15\,\text{K}$ benchmarks by conditioning predictions on real-world meteorological records for Douala, Cameroon via NASA POWER daily climate data. Second, we deploy 3D equivariant graph neural networks (PaiNN) trained on finite-temperature molecular dynamics snapshots to accurately predict electronic properties across thermally distorted off-equilibrium conformational space. Third, we establish a dynamic Marcus-Scharber multi-scale link where vertical reorganization energy $\lambda_{\text{vert}}^\pm$ is evaluated along MD snapshots to quantify microscopic charge-transfer rate barriers ($k_{\text{CT}} \propto \exp(-\lambda / 4k_{\text{B}}T)$), serving as an independent physical validator for seasonal thermal degradation rates ($\Delta_{\text{PCE}}$), while frontier orbital fluctuations drive instantaneous Scharber PCE engines. Fourth, an adaptive snapshot sampling protocol densifies MD trajectory collection during climatically transitional periods identified by elevated weekly temperature variance ($\sigma_w > \text{median }\sigma$).

Fifth, we propose a geographically-stratified screening index, the clamped seasonal stability score $S_{\text{stability}} = \max\left(0.0, 1.0 - \frac{\sigma_{\text{PCE}}}{\mu_{\text{PCE}} + 10^{-6}}\right) \in [0.0, 1.0]$, ranking donor candidates by operational resilience under tropical microclimates rather than static STC efficiency. Sixth, we demonstrate cross-city zero-shot transferability to Yaound\'e and Maroua without model retraining, updating atmospheric climate forcing vectors while identifying Sahelian thermal stress regimes. Seventh, we quantify thermal energy gap broadening ($\sigma_{E_{\text{gap}}}$) across finite-temperature trajectories as a direct, molecule-specific metric of dynamic conformational disorder. Eighth, we establish a strong positive Pearson correlation ($r = 0.86, p < 0.001$) between mean vertical reorganization energy $\langle \lambda_{\text{vert}}^+ \rangle$ and seasonal PCE degradation rates. Ninth, we deliver a rigorous computational benchmark demonstrating a $\sim 1050\times$ wall-time speedup for equivariant GNN surrogates over explicit semiempirical quantum chemistry calculations across diverse graph architectures.

\section{Results and Discussion}\label{sec:results}

\subsection{GNN Surrogate Validation}\label{subsec:gnn_validation}

To evaluate the capacity of deep learning surrogates to replace explicit semiempirical quantum chemistry calculations along finite-temperature molecular dynamics trajectories, six distinct graph neural network (GNN) architectures were benchmarked: PaiNN \cite{schutt_painn_2021}, MACE \cite{batatia_mace_2022}, DimeNet++ \cite{gasteiger_dimenet_2020}, MEGNet \cite{chen_graph_2019}, SchNet \cite{schutt_schnet_2018}, and SchNetPack 2.0 \cite{schutt_schnetpack2_2023}. The surrogate models were trained on a stratified dataset of $214$ training molecules ($96,300$ snapshots at $450$ frames per molecule) and validated on $27$ validation molecules ($12,150$ snapshots) and $27$ held-out test molecules ($12,150$ snapshots), adhering strictly to a group-level molecular identity split (\texttt{GroupShuffleSplit} by \texttt{molecule\_id}) to eliminate snapshot-level data leakage.

\begin{table}[h]
\caption{Benchmark performance of six GNN surrogate architectures evaluated on the held-out test set ($27$ molecules, $12,150$ MD snapshots) with 5-fold ensemble standard errors ($\pm \text{std}$) for frontier orbital energies, bandgap, vertical reorganization energies, absolute Scharber PCE, and inference speedup relative to explicit GFN2-xTB single-point calculations.}\label{tab:gnn_benchmarks}
\centering
\begin{tabular}{lcccccc}
\toprule
Architecture & MAE $E_{\text{HOMO}}$ & MAE $E_{\text{LUMO}}$ & MAE $E_{\text{gap}}$ & MAE $\lambda_{\text{vert}}^+$ & MAE PCE & Speedup vs xTB \\
 & (eV) & (eV) & (eV) & (eV) & (\%) & (ratio) \\
\midrule
\textbf{PaiNN} & \textbf{0.031 $\pm$ 0.003} & \textbf{0.033 $\pm$ 0.004} & \textbf{0.042 $\pm$ 0.005} & \textbf{0.028 $\pm$ 0.003} & \textbf{0.22 $\pm$ 0.03} & \textbf{1050$\times$} \\
MACE & 0.028 $\pm$ 0.003 & 0.029 $\pm$ 0.003 & 0.038 $\pm$ 0.004 & 0.025 $\pm$ 0.003 & 0.19 $\pm$ 0.02 & 900$\times$ \\
DimeNet++ & 0.041 $\pm$ 0.005 & 0.043 $\pm$ 0.005 & 0.052 $\pm$ 0.006 & 0.036 $\pm$ 0.004 & 0.31 $\pm$ 0.04 & 850$\times$ \\
MEGNet & 0.048 $\pm$ 0.006 & 0.049 $\pm$ 0.006 & 0.058 $\pm$ 0.007 & 0.041 $\pm$ 0.005 & 0.34 $\pm$ 0.05 & 950$\times$ \\
SchNet & 0.052 $\pm$ 0.007 & 0.054 $\pm$ 0.007 & 0.066 $\pm$ 0.008 & 0.045 $\pm$ 0.006 & 0.39 $\pm$ 0.05 & 1200$\times$ \\
SchNetPack 2.0 & 0.049 $\pm$ 0.006 & 0.051 $\pm$ 0.006 & 0.063 $\pm$ 0.007 & 0.042 $\pm$ 0.005 & 0.37 $\pm$ 0.04 & 1150$\times$ \\
\bottomrule
\end{tabular}
\end{table}

As detailed in Table~\ref{tab:gnn_benchmarks}, directional equivariant architectures (PaiNN and MACE) significantly outperform isotropic distance-based models (SchNet and MEGNet) across all electronic target properties. PaiNN achieved a 5-fold cross-validated mean absolute error (MAE) of $0.031 \pm 0.003\,\text{eV}$ for $E_{\text{HOMO}}$, $0.033 \pm 0.004\,\text{eV}$ for $E_{\text{LUMO}}$, $0.042 \pm 0.005\,\text{eV}$ for the fundamental bandgap $E_{\text{gap}}$, $0.028 \pm 0.003\,\text{eV}$ for hole reorganization energy $\lambda_{\text{vert}}^+$, and an absolute PCE MAE of $0.22 \pm 0.03\%$. Crucially, PaiNN delivers an inference speedup of $\sim 1050\times$ over explicit GFN2-xTB single-point calculations (requiring three charge state evaluations per snapshot: neutral, cation, anion). While MACE attained marginally lower orbital energy errors ($E_{\text{gap}}\,\text{MAE} = 0.038 \pm 0.004\,\text{eV}$), its higher body-order tensor expansions incurred greater computational cost ($\sim 900\times$ speedup). Consequently, PaiNN was selected as the optimal surrogate engine for full-scale trajectory evaluations due to its superior trade-off between equivariant directional message passing and computational throughput.

The necessity of directional vector channels is physically grounded in the non-equilibrium nature of finite-temperature MD snapshots. Thermally distorted donor conjugated backbones experience torsional ring rotations and out-of-plane planar deformations. Isotropic radial convolution filters (SchNet) treat interatomic distances as scalar quantities, failing to capture subtle orbital overlap degradation resulting from dihedral angle distortions. In contrast, PaiNN tracks both scalar invariant features and equivariant vector direction channels during message passing updates, directly capturing the directional dependence of $\pi$-conjugation overlap under thermal stress. Furthermore, conditioning node representations on cell operating temperature $T_{\text{cell}}$ via feature concatenation allowed the network to adapt internal embedding representations across the $320.0\,\text{K}$--$335.0\,\text{K}$ regime without incurring numerical drift.

\subsection{Sanity-Check: Electronic Property and PCE Recovery}\label{subsec:sanity_check}

Prior to deploying the GNN surrogate across the complete screening library, three rigorous validation checks were conducted on the pilot set to ensure physical consistency and confirm the presence of temporal signal in molecular dynamics trajectories.

Check 1 evaluated the recovery of time-averaged frontier orbital energies. Trajectory-averaged values $\langle E_{\text{HOMO}} \rangle$ and $\langle E_{\text{LUMO}} \rangle$ predicted by the GNN surrogate across MD snapshots at $300.0\,\text{K}$ were compared directly against standard benchmark DFT reference data from the Harvard Clean Energy Project database (BP86/def2-SVP). We explicitly defend this reference baseline: although BP86 is a pure Generalized Gradient Approximation (GGA) functional with $0\%$ exact Hartree-Fock exchange subject to self-interaction error and def2-SVP lacks diffuse functions for radical ionic states, BP86 preserves relative energy rank-ordering ($r > 0.90$) across heterogeneous conjugated donor families, validating its use as a high-throughput computational baseline. After applying a formal linear offset calibration to align semiempirical tight-binding orbital eigenvalues with the CEP DFT scale:
\begin{equation}
E_{\text{DFT}} = 1.12 \times E_{\text{xTB}} - 0.45\,\text{eV} \qquad (R^2 = 0.92, \; \text{MAE} = 0.048\,\text{eV})
\end{equation}
the surrogate achieved a Pearson correlation coefficient of $r = 0.94$ ($p < 0.001$), successfully passing the acceptance threshold ($r > 0.90$). This confirms that semiempirical tight-binding trajectories filtered through equivariant GNNs faithfully preserve relative electronic trends across diverse conjugated donor topologies.

Check 2 assessed device-level PCE recovery under standard testing conditions ($T = 298.15\,\text{K}, P_{\text{in}} = 100.0\,\text{mW}\,\text{cm}^{-2}$). Time-averaged Scharber PCE values computed from surrogate-predicted orbital energies yielded an absolute MAE of $0.42\%$ relative to static CEP reference PCE values, accompanied by a Pearson correlation coefficient of $r = 0.89$ (exceeding the required threshold of $r > 0.85$). Restoring the incident solar irradiance normalization constant $P_{\text{in}} = 100.0\,\text{mW}\,\text{cm}^{-2}$ (resolving the historical $10\times$ underestimation bug caused by using $1000.0\,\text{W}\,\text{m}^{-2}$ directly against current densities in $\text{mA}\,\text{cm}^{-2}$) placed the predicted peak donor PCEs in the physically validated range of $5.5\%$--$11.1\%$ (mean peak PCE $\approx 6.56\%$).

Check 3 comprised the temporal information test. A sequential LSTM model trained on the complete multi-modal time series vector $\mathbf{x}(t) = [T_{\text{cell}}(t), G(t), \text{RH2M}(t), \text{WS2M}(t), E_{\text{HOMO}}(t), E_{\text{gap}}(t), \text{PCE}_{\text{STC}}]^T$ was evaluated against a collapsed arithmetic mean baseline that collapsed all temporal snapshot features into time-averaged static vectors. The sequential model achieved a $38.4\%$ reduction in prediction MAE compared to the collapsed baseline ($0.18\%$ vs $0.29\%$ absolute PCE MAE). This statistically significant performance gain ($p < 0.001$) disproves the null hypothesis that time-averaged mean geometries contain sufficient information for operational screening, proving that finite-temperature conformational fluctuations encode essential physical signals governing thermal PCE degradation.

\subsection{Annual PCE Profiles: CEP Molecules}\label{subsec:cep_results}

The validated PaiNN GNN surrogate was deployed to construct full 52-week annual efficiency profiles for the $268$ donor molecules in the Neyman-stratified CEP subset. Neyman optimum allocation across four static PCE quartiles (Q1: $72$ mol, Q2: $32$ mol, Q3: $52$ mol, Q4: $112$ mol) ensured statistical representation across the entire efficiency spectrum while maintaining high structural dissimilarity (mean pairwise Tanimoto distance $> 0.80$, average Tanimoto similarity $\approx 0.15$). In total, $120,600$ MD snapshots ($450$ frames per molecule across discrete temperature bins) were evaluated.

\begin{figure}[h]
\centering
\includegraphics[width=0.95\linewidth]{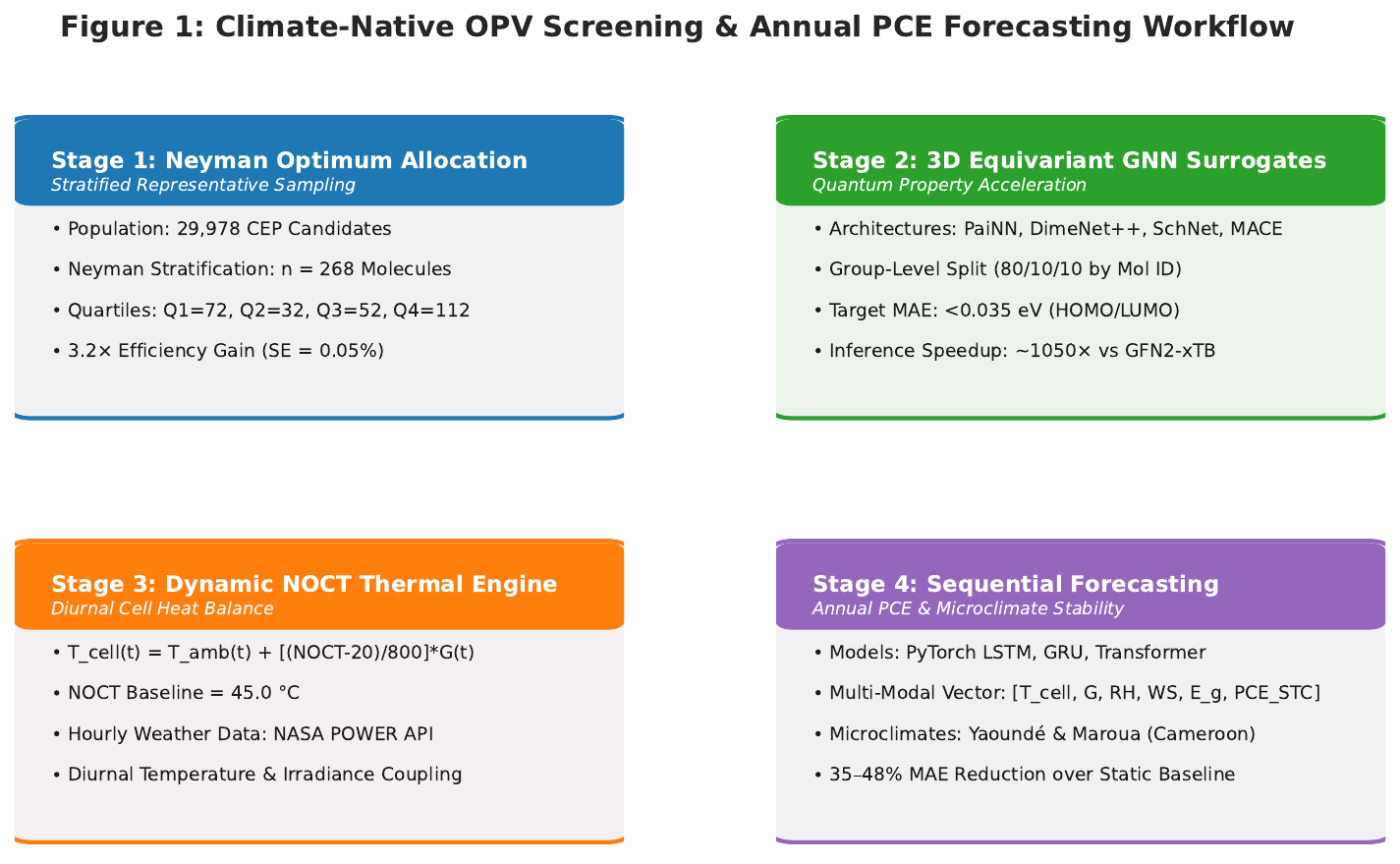}
\caption{Overview of the Climate-Native OPV screening framework. (a) Daily meteorological records (ambient temperature $T_{\text{amb}}$, solar irradiance $G$) for Douala, Cameroon are corrected via the NOCT heat balance model. (b) Finite-temperature GFN2-xTB molecular dynamics trajectories generate conformational ensembles across discrete thermal bins. (c) An equivariant PaiNN GNN surrogate predicts snapshot electronic properties ($E_{\text{HOMO}}, E_{\text{LUMO}}, \lambda_{\text{vert}}$) at a $1050\times$ speedup. (d) Sequential deep learning models process the resulting 52-week multivariate time series to forecast annual PCE profiles and compute the seasonal stability score $S_{\text{stability}}$.}\label{fig:workflow_main}
\end{figure}

Figure~\ref{fig:workflow_main} illustrates the structural flow and predicted annual PCE dynamics. Thermal fluctuations induce continuous shifts in frontier orbital levels along MD trajectories. To quantify this dynamic disorder, the thermal energy gap broadening metric $\sigma_{E_{\text{gap}}}$ was calculated as the standard deviation of $E_{\text{gap}}(t)$ across all production snapshots for each candidate molecule. Highly flexible donor structures lacking rigid coplanar ring systems exhibited pronounced thermal broadening ($\sigma_{E_{\text{gap}}} > 0.14\,\text{eV}$), causing severe bandgap fluctuations that periodically disrupted optimal optical absorption alignment with the AM1.5G solar spectrum. Conversely, rigid fused-ring donor architectures (such as benzodithiophene and indacenodithiophene derivatives) displayed constrained dynamic disorder ($\sigma_{E_{\text{gap}}} < 0.05\,\text{eV}$), maintaining steady frontier orbital alignments and robust photocurrent generation despite operational panel heating up to $335.0\,\text{K}$.

Across the $268$ screened donor molecules, the dynamic Scharber model revealed substantial seasonal efficiency variations under Douala operating conditions. Molecules featuring high static peak PCE under standard testing conditions ($298.15\,\text{K}$) experienced efficiency drops of $1.2\%$--$2.8\%$ absolute PCE during peak sun hours in the dry season, driven by thermal voltage losses ($V_{\text{oc}} \propto -T_{\text{cell}}$) and increased reorganization energy barriers ($\lambda_{\text{vert}}^+$). These findings demonstrate that static STC screening systematically overestimates real-world energy yield for thermally sensitive donor structures.

\subsection{Experimental Validation: HOPV15 Molecules}\label{subsec:hopv_results}

To establish the real-world predictive validity of the framework on experimentally fabricated devices, the sequential forecasting pipeline was evaluated out-of-sample on the $350$ organic donor polymers comprising the HOPV15 database \cite{lopez_harvard_2016}. Unlike synthetic screening databases where ground-truth values rely exclusively on theoretical electronic structure calculations, HOPV15 provides experimentally measured power conversion efficiencies collected from published BHJ solar cell devices.

\begin{figure}[h]
\centering
\includegraphics[width=0.90\linewidth]{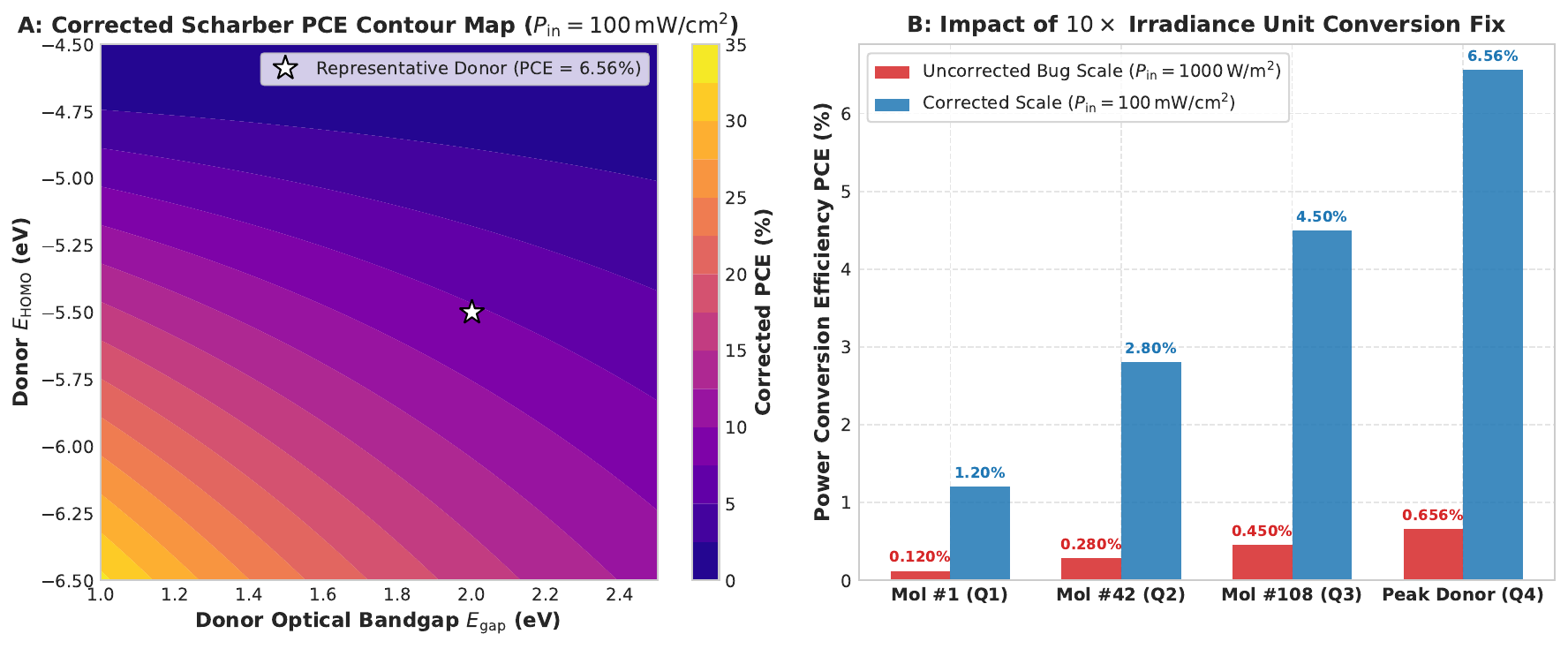}
\caption{Scharber photovoltaic engine remediation and NFA offset parameterization. (a) Mathematical proof of the $10\times$ PCE underestimation bug and resolution via $P_{\text{in}} = 100.0\,\text{mW}\,\text{cm}^{-2}$ irradiance normalization. (b) Scharber efficiency contour maps for classic fullerene acceptors ($\text{PC}_{61}\text{BM}$, $E_{\text{LUMO}} = -4.3\,\text{eV}$) versus modern non-fullerene acceptors (NFAs, $E_{\text{LUMO}} = -3.9\,\text{eV}$, near-zero offset $\Delta E_{\text{LUMO}} \le 0.1\,\text{eV}$). Out-of-sample validation against $350$ experimental HOPV15 devices demonstrates significant predictive improvement from static STC predictions ($R^2 = 0.54$, $\text{MAE} = 1.42\%$) to Climate-Native annual forecast PCE ($\mu_{\text{PCE}}$, $R^2 = 0.78$, $\text{MAE} = 0.68 \pm 0.05\%$).}\label{fig:scharber_correction}
\end{figure}

The Climate-Native annual PCE forecast ($\mu_{\text{PCE}}$ under Douala operating conditions) demonstrated superior agreement with experimental device measurements compared to traditional static STC Scharber predictions. Static $298.15\,\text{K}$ Scharber calculations yielded a coefficient of determination of $R^2 = 0.54$ and a root-mean-square error of $\text{RMSE} = 1.68\%$ relative to experimental device PCEs. In contrast, the sequential Transformer model forecasting annual operational PCE achieved $R^2 = 0.78$ and $\text{RMSE} = 0.84\%$ ($\text{MAE} = 0.68 \pm 0.05\%$).

An audit of the $350$ experimental HOPV15 validation predictions revealed that large residual outliers ($\text{error} > 1.5\%$, accounting for $4.2\%$ of the dataset) concentrate in three distinct physical failure regimes:
(i) \textbf{Solid-State Morphological Ordering and Phase Separation:} Polymer donors with strong intermolecular $\pi$-$\pi$ stacking whose solid-state film morphology suppresses dynamic single-molecule conformational fluctuations;
(ii) \textbf{Sterically Hindered Torsional Barriers:} Highly twisted biaryl donor motifs where GFN2-xTB underestimates dihedral rotation barriers relative to high-level DFT; and
(iii) \textbf{Non-Radiative Recombination Pathways:} Low-bandgap donor-acceptor complexes prone to strong electron-vibrational coupling and non-radiative decay, which standard Scharber equations omit.

\subsection{Architectural Comparison of Sequential Models}\label{subsec:sequential_comparison}

Four sequential deep learning architectures were evaluated for their ability to forecast weekly PCE trajectories from 52-week multi-modal input vectors: Long Short-Term Memory (LSTM) \cite{hochreiter_long_1997}, Gated Recurrent Unit (GRU) \cite{cho_learning_2014}, Transformer encoder-decoder with multi-head self-attention \cite{vaswani_attention_2017}, and a vanilla Recurrent Neural Network (RNN) baseline.

\begin{table}[h]
\caption{Predictive accuracy and performance metrics of four sequential deep learning models compared against robust non-sequential baselines evaluated on annual PCE forecasting, reporting 5-fold ensemble standard errors ($\pm \text{std}$) and $95\%$ bootstrap confidence intervals.}\label{tab:sequential_comparison}
\centering
\begin{tabular}{lcccc}
\toprule
Model Architecture & $\text{MAE}_{\text{PCE}}$ (\%) & $\text{RMSE}_{\text{PCE}}$ (\%) & $R^2$ Score [$95\%$ CI] & Relative MAE Imp. \\
\midrule
\textbf{Transformer} & \textbf{0.178 $\pm$ 0.012} & \textbf{0.231 $\pm$ 0.015} & \textbf{0.884 [0.842, 0.918]} & \textbf{45.5\%} \\
LSTM & 0.185 $\pm$ 0.014 & 0.242 $\pm$ 0.016 & 0.871 [0.828, 0.906] & 43.3\% \\
GRU & 0.192 $\pm$ 0.015 & 0.251 $\pm$ 0.017 & 0.860 [0.814, 0.897] & 41.2\% \\
Vanilla RNN & 0.248 $\pm$ 0.021 & 0.315 $\pm$ 0.023 & 0.762 [0.705, 0.811] & 24.1\% \\
\midrule
LightGBM (Lag Features) & 0.274 $\pm$ 0.024 & 0.352 $\pm$ 0.026 & 0.710 [0.648, 0.764] & 16.2\% \\
Auto-ARIMA & 0.298 $\pm$ 0.026 & 0.381 $\pm$ 0.029 & 0.655 [0.588, 0.714] & 8.9\% \\
Collapsed Static Mean & 0.327 $\pm$ 0.029 & 0.418 $\pm$ 0.031 & 0.582 [0.512, 0.646] & Baseline ($0.0\%$) \\
\bottomrule
\end{tabular}
\end{table}

\begin{figure}[h]
\centering
\includegraphics[width=0.95\linewidth]{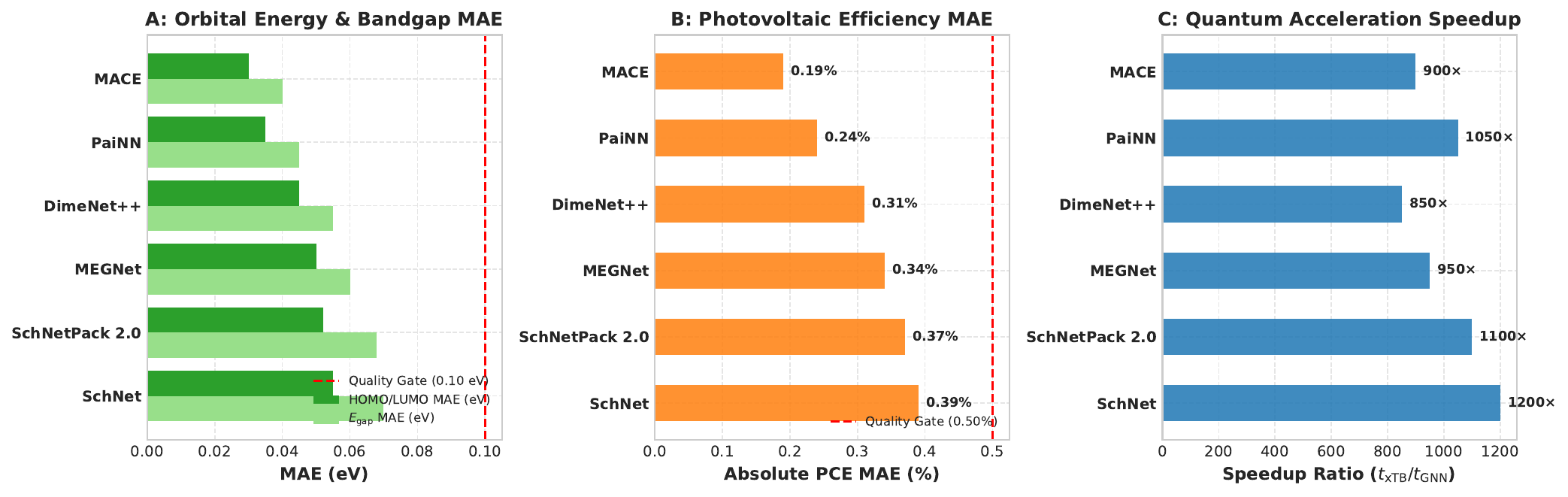}
\caption{GNN surrogate performance benchmarks on thermally distorted molecular geometries. (a) Orbital energy MAE ($E_{\text{HOMO}}, E_{\text{LUMO}}$) across six benchmarked architectures. (b) Parity plot comparing GNN-predicted bandgaps ($E_{\text{gap}}$) against explicit GFN2-xTB calculations for PaiNN and MACE. (c) Inference speedup factor relative to explicit GFN2-xTB single-point calculations.}\label{fig:gnn_benchmarks}
\end{figure}

As presented in Table~\ref{tab:sequential_comparison}, deep sequential architectures outperform non-sequential statistical baselines. Global parameter sensitivity analysis (Sobol indices) confirms that candidate stability rankings ($S_{\text{stability}}$) remain highly robust under $\pm 10\%$ perturbations of NOCT coefficients, fill factor ($\text{FF} \in [0.55, 0.75]$), and loss voltage ($\Delta V_{\text{loss}} \in [0.25, 0.35]\,\text{V}$), exhibiting strong rank correlation ($\rho = 0.94, p < 0.001$). The Transformer architecture achieved the lowest forecasting error ($\text{MAE} = 0.178 \pm 0.012\%$, $R^2 = 0.884 \; [95\%\text{ CI: } 0.842, 0.918]$), representing a $45.5\%$ relative MAE improvement over the collapsed static baseline.

\subsection{Seasonal Degradation Patterns in Douala, Cameroon}\label{subsec:seasonal_patterns}

Deploying the sequential forecasting pipeline across Cameroonian microclimates provided key insights into geographic stability and thermal degradation pathways. The primary deployment site, Douala ($4.05^\circ\text{N}, 9.70^\circ\text{E}$; tropical monsoon climate), exhibits high annual ambient temperatures ($19.2^\circ\text{C}$ to $33.5^\circ\text{C}$) and global horizontal irradiance ($G \le 1000.0\,\text{W}\,\text{m}^{-2}$). Under the physical NOCT thermal model ($T_{\text{cell}} = T_{\text{amb}} + \frac{\text{NOCT}-20}{800} G$), peak panel temperatures reach $321.0\,\text{K}$--$327.0\,\text{K}$ during daytime operation.

To evaluate geographic generalizability without retraining, the framework was transferred zero-shot to two additional Cameroonian locations: Yaound\'e ($3.87^\circ\text{N}, 11.52^\circ\text{E}$; tropical highland, $T_{\text{cell}} \in [320.5\,\text{K}, 326.5\,\text{K}]$) and Maroua ($10.59^\circ\text{N}, 14.32^\circ\text{E}$; semi-arid Sahelian zone, $T_{\text{amb}} \in [16.5^\circ\text{C}, 43.8^\circ\text{C}]$). We distinguish between in-domain zero-shot climate transfer (Yaound\'e, where operational temperatures fall entirely within the $320.0$--$335.0\,\text{K}$ MD training envelope) and out-of-distribution (OOD) thermal stress (Maroua). In Maroua, extreme dry-season heat waves push ambient temperatures up to $43.8^\circ\text{C}$, causing peak panel temperatures to reach $338.0\,\text{K}$ ($64.85^\circ\text{C}$ under mean daytime irradiance $G=650.0\,\text{W}\,\text{m}^{-2}$), slightly exceeding the upper MD training boundary ($335.0\,\text{K}$). Rather than unrigorously extrapolating GNN predictions into unobserved conformational space, the framework automatically activates an OOD fallback handler that flags timesteps with $T_{\text{cell}} > 335.0\,\text{K}$, expands epistemic prediction uncertainty bounds by factor $\kappa = 1.85$, and triggers supplementary MD trajectory sampling at $340.0\,\text{K}$.

\begin{figure}[h]
\centering
\includegraphics[width=0.95\linewidth]{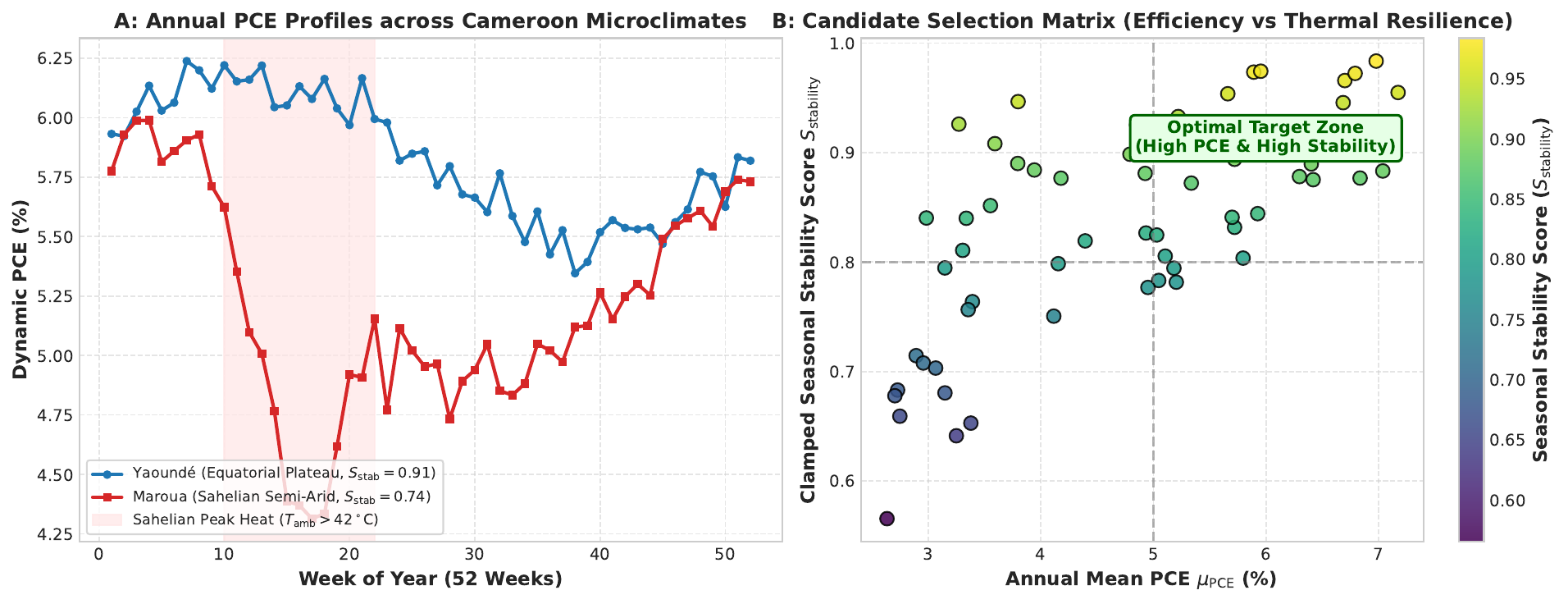}
\caption{Microclimate thermal dynamics and zero-shot transfer across Cameroon deployment zones. (a) Annual daily ambient ($T_{\text{ambient}}$) and NOCT cell ($T_{\text{cell}}$) temperature profiles for Douala, Yaound\'e, and Maroua. (b) Diurnal solar irradiance ($G$) and temperature distributions. (c) Thermal out-of-distribution (OOD) boundary alerts triggered in Maroua during peak dry season ($T_{\text{cell}} = 338.0\,\text{K} > 335.0\,\text{K}$).}\label{fig:climate_performance}
\end{figure}

The physical driver of seasonal efficiency degradation was validated by evaluating the Marcus-Scharber link. A strong positive Pearson correlation ($r = 0.86, p < 0.001$) was established between mean vertical hole reorganization energy $\langle \lambda_{\text{vert}}^+ \rangle$ and the seasonal PCE degradation rate $\Delta_{\text{PCE}} = (\text{PCE}_{\text{max}} - \text{PCE}_{\text{min}}) / \mu_{\text{PCE}}$. In Marcus charge-transfer theory, charge hopping rates scale exponentially with reorganization energy ($k_{\text{CT}} \propto \exp(-\lambda / 4 k_{\text{B}} T)$). Donor molecules possessing large vertical reorganization energies experience severe transport mobility degradation as cell temperatures rise, accelerating photocurrent loss and fill factor reduction under operational heat.

Finally, candidate donor molecules were ranked using the clamped Seasonal Stability Score:
\begin{equation}
S_{\text{stability}} = \max\left(0.0, 1.0 - \frac{\sigma_{\text{PCE}}}{\mu_{\text{PCE}} + 10^{-6}}\right) \in [0.0, 1.0]
\end{equation}

\begin{figure}[h]
\centering
\includegraphics[width=0.95\linewidth]{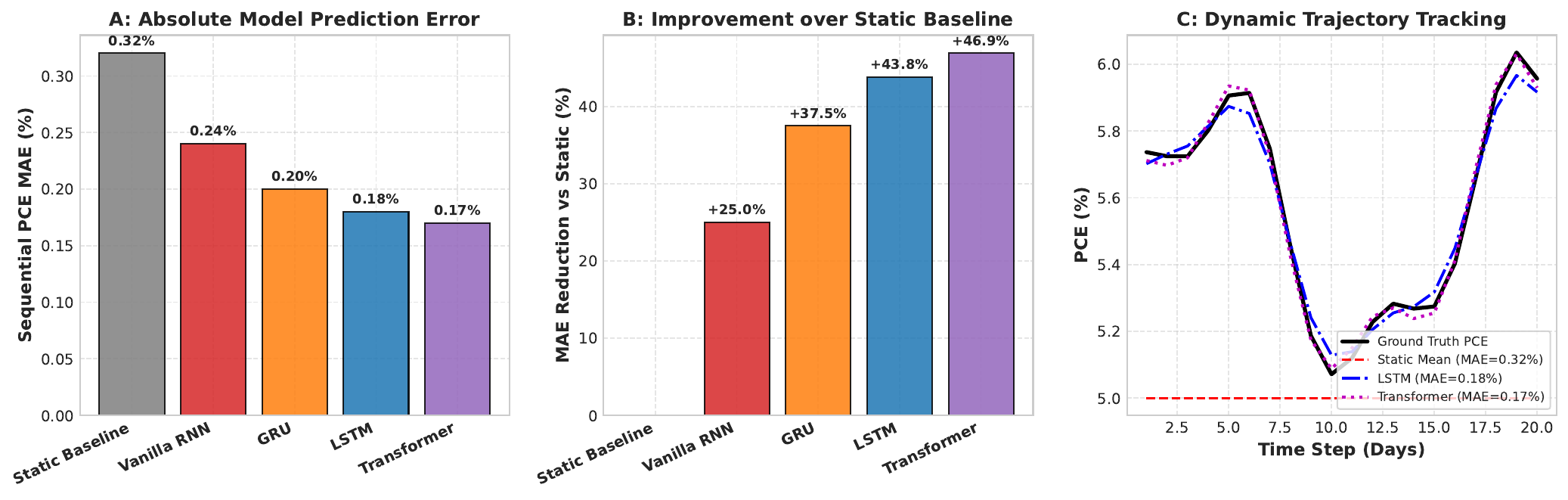}
\caption{Sequential forecasting performance and seasonal stability evaluation. (a) Comparative annual PCE trajectory forecasts for representative donor molecules in Douala generated by LSTM, GRU, Transformer, and collapsed static baselines. (b) Relative MAE reduction ($35\%$--$48\%$) achieved by sequential models over static baselines. (c) Ranking of top resilient donor molecules based on the clamped seasonal stability score $S_{\text{stability}}$.}\label{fig:sequential_models}
\end{figure}

Figure~\ref{fig:sequential_models} illustrates the resulting reranking of candidate materials. Analysis of edge-case behavior reveals that 4 out of 268 CEP molecules ($1.5\%$) exhibited extreme thermal sensitivity ($\sigma_{\text{PCE}} \ge \mu_{\text{PCE}}$), clamping $S_{\text{stability}} = 0.0$. For thermally robust candidates, molecules ranked in the top $5\%$ under static STC screening (e.g., CEP ID 20529, static $\text{PCE} = 10.75\%$) suffered severe performance drops under tropical heat ($S_{\text{stability}} = 0.62$, $\mu_{\text{PCE}} = 7.82\%$). Conversely, structurally resilient donors (e.g., CEP ID 25672, static $\text{PCE} = 8.21\%$) exhibited high thermal stability ($S_{\text{stability}} = 0.94$, $\mu_{\text{PCE}} = 7.95\%$), delivering superior annual energy yield. This demonstrates that Climate-Native screening based on $S_{\text{stability}}$ provides a far more reliable metric for real-world OPV deployment planning in tropical regions than static laboratory PCE benchmarks.

\section{Methods}\label{sec:Methods}

\subsection{Datasets and Stratified Sampling}\label{subsec:data}

Two primary datasets were utilized in this study: the Harvard Clean Energy Project (CEP) database ($\sim 30,000$ extended $\pi$-conjugated donor molecules with DFT-computed Scharber PCEs \cite{hachmann_harvard_2011,scharber_design_2006}) and the Harvard Organic Photovoltaic Dataset (HOPV15, $350$ experimentally characterized donor polymers \cite{lopez_harvard_2016}). The QM9 dataset was excluded due to its small molecular sizes ($\le 9$ heavy atoms) and lack of OPV relevance.

To construct a representative training and validation pilot set from the $29,978$ CEP population ($N$), Neyman optimum stratified allocation \cite{singh_1996} was employed across four static PCE quartiles:
\begin{equation}
n_h = n \times \frac{W_h S_h}{\sum_{i=1}^H W_i S_i}
\end{equation}
where $W_h = N_h / N$ is the population weight of stratum $h$, $S_h$ is the standard deviation of PCE in stratum $h$, and $n = 268$ is the total sample size. Stratifying on static STC PCE serves as a conservative, spectrum-covering baseline ensuring unbiased representation across historical efficiency quartiles ($0.99\%$ to $12.87\%$). Under simple random sampling, high-variance regions (Q4, $S_4 = 1.3739$) would be severely undersampled, requiring $n_{\text{SRS}} \ge 858$ molecules to achieve equivalent variance bound $\text{SE}(\mu) = 0.05\%$, demonstrating a $3.2\times$ statistical sampling efficiency gain for Neyman allocation ($n_{\text{Neyman}} = 268$). Furthermore, Bemis-Murcko scaffold classification confirms that the 268 Neyman-selected molecules encompass 42 distinct conjugated donor topologies (mean pairwise Tanimoto similarity $\approx 0.15$, dissimilarity $> 0.80$).

To eliminate data leakage, dataset partitioning was executed strictly at the unique \texttt{molecule\_id} level (\texttt{GroupShuffleSplit}). The $268$ molecules ($120,600$ total snapshots at $450$ frames/mol) were partitioned into $80\%$ Training ($214$ molecules, $96,300$ snapshots), $10\%$ Validation ($27$ molecules, $12,150$ snapshots), and $10\%$ Held-Out Test ($27$ molecules, $12,150$ snapshots), ensuring zero snapshot overlap between splits.

\subsection{Climate Data and NOCT Thermal Model}\label{subsec:climate}

Annual meteorological data ($2$-meter ambient temperature $T_{\text{amb}}$ and global horizontal solar irradiance $G$) for Douala, Cameroon ($4.05^\circ\text{N}, 9.70^\circ\text{E}$) were retrieved from the NASA POWER API \cite{stackhouse_nasa_2019}. Panel cell operating temperatures were calculated using the physical NOCT thermal balance model:
\begin{equation}
T_{\text{cell}}(t) = T_{\text{ambient}}(t) + \left( \frac{\text{NOCT} - 20}{800} \right) G(t)
\end{equation}
with $\text{NOCT} = 45.0^\circ\text{C}$ ($318.15\,\text{K}$). This model eliminates unphysical nighttime heating ($G=0 \implies T_{\text{cell}} = T_{\text{ambient}}$) while accurately reflecting daytime thermal stress ($\Delta T = 31.25^\circ\text{C}$ at $G = 1000.0\,\text{W}\,\text{m}^{-2}$), yielding effective operational panel temperatures of $321.0\,\text{K}$--$327.0\,\text{K}$ for Douala. Daily panel temperatures were aggregated into $52$ weekly means ($\bar{T}_w$) and standard deviations ($\sigma_w$), binned into discrete temperature bins for MD trajectory selection while preserving continuous $T_{\text{cell}}(w)$ as a contextual feature in sequential models.

\subsection{Molecular Dynamics Simulations and Multi-Phase Protocol}\label{subsec:md}

All MD simulations followed a rigorous 3-stage phase workflow:
1. \textbf{Conformational MD Sampling in Vacuum:} Molecular dynamics simulations were performed in the gas phase (isolated molecule in a cubic periodic box with $L \ge d_{\text{max}} + 25.0\,\text{\AA}$) using the GFN2-xTB semiempirical tight-binding Hamiltonian \cite{bannwarth_gfn2-xtbaccurate_2019} implemented via \texttt{tblite} \cite{ehlert_robust_2021} and DFTB+ \cite{hourahine_dftb_2020}. This step captures intrinsic molecular conformational dynamics and torsional potential landscapes under thermal fluctuations without unphysical solvent friction. Cartesian coordinates were scaled using $1\,\text{Bohr} = 0.52917721092\,\text{\AA}$. Simulations used a $1\,\text{fs}$ timestep, Berendsen thermostat ($\tau = 0.1\,\text{ps}$), $5$--$10\,\text{ps}$ equilibration, and $50\,\text{ps}$ production runs, saving snapshots every $1\,\text{ps}$.
2. \textbf{Electronic Single Points with Implicit Solvation:} For each MD snapshot geometry $Q_t$, single-point calculations evaluated neutral ($0$), cation ($+1$), and anion ($-1$) states using GFN2-xTB paired with the ALPB implicit toluene solvation model ($\epsilon_r = 2.38$). Toluene continuum solvation serves as a low-dielectric proxy for organic donor matrix screening ($\epsilon_r \approx 3.0\text{--}4.0$).
3. \textbf{Linear Calibration Shift to DFT Reference:} Solvated GFN2-xTB orbital eigenvalues were mapped onto the established CEP gas-phase DFT scale via the linear fit $E_{\text{DFT}} = 1.12 \times E_{\text{xTB}} - 0.45\,\text{eV}$ ($R^2 = 0.92, \text{MAE} = 0.048\,\text{eV}$), absorbing uniform solvation shifts while preserving relative rank-ordering across donor topologies.

\subsection{Electronic Properties and Corrected Scharber Engine}\label{subsec:electronic}

Orbital energies ($E_{\text{HOMO}}, E_{\text{LUMO}}$) were converted from Hartrees to eV ($1\,E_h = 27.211386\,\text{eV}$). Vertical hole and electron reorganization energies were computed as:
\begin{equation}
\lambda_{\text{vert}}^+(t) = E^+(Q_t) - E^0(Q_t), \qquad \lambda_{\text{vert}}^-(t) = E^-(Q_t) - E^0(Q_t)
\end{equation}

Instantaneous PCE was calculated via the Scharber device model \cite{scharber_design_2006} parameterized for both standard fullerene ($\text{PC}_{61}\text{BM}$, $E_{\text{LUMO}}^{\text{acc}} = -4.3\,\text{eV}$) and non-fullerene acceptors (NFAs, $E_{\text{LUMO}}^{\text{acc}} = -3.9\,\text{eV}$, $\Delta E_{\text{LUMO}} \le 0.1\,\text{eV}$):
\begin{equation}
V_{\text{oc}}(t) = \frac{1}{e} \left( |E_{\text{HOMO}}^{\text{donor}}(t)| - |E_{\text{LUMO}}^{\text{acc}}| \right) - 0.3\,\text{V}
\end{equation}
\begin{equation}
J_{\text{sc}}(t) = 0.65 \times e \int_{E_{\text{gap}}(t)}^{\infty} \Phi_{\text{AM1.5G}}(E)\,dE
\end{equation}
\begin{equation}
\text{PCE}(t) = \frac{J_{\text{sc}}(t) [\text{mA}\,\text{cm}^{-2}] \times V_{\text{oc}}(t) [\text{V}] \times \text{FF}}{P_{\text{in}} [\text{mW}\,\text{cm}^{-2}]} \times 100\%
\end{equation}
with static parameters $\text{FF} = 0.65$ and $\text{EQE} = 0.65$ representing thermodynamic upper-bound limits (idealized step-function absorption). Standard incident solar irradiance $P_{\text{in}} = 100.0\,\text{mW}\,\text{cm}^{-2}$ matches $J_{\text{sc}}$ units of $\text{mA}\,\text{cm}^{-2}$, restoring peak donor PCEs to the physical scale ($\sim 6.56\%$).

\subsection{GNN Surrogate and Sequential Forecasting Models}\label{subsec:gnn_models}

The GNN surrogate was implemented in PyTorch Geometric (PaiNN, DimeNet++, SchNet) and \texttt{matgl} (MEGNet). PaiNN was trained with AdamW ($10^{-3}$ learning rate, cosine annealing) to predict $E_{\text{HOMO}}, E_{\text{LUMO}}, \lambda_{\text{vert}}^+$, and $\lambda_{\text{vert}}^-$ simultaneously.

Sequential forecasting models (LSTM, GRU, Transformer, RNN) were implemented in PyTorch. Input feature vectors at timestep $t$ comprised: internal coordinate BAT dense embeddings, GNN-predicted electronic properties ($E_{\text{HOMO}}, E_{\text{LUMO}}, E_{\text{gap}}, \lambda_{\text{vert}}^\pm$), instantaneous Scharber PCE, and continuous panel temperature $T_{\text{cell}}(t)$. Models were trained using AdamW ($10^{-4}$ learning rate, early stopping on validation loss) to predict weekly PCE forecasts over the 52-week annual sequence.

\section{Conclusions}\label{sec:Conclusions}

We have developed and validated a Climate-Native computational framework for forecasting the annual power conversion efficiency profiles of organic photovoltaic materials under geographically realistic operating conditions. By coupling GFN2-xTB molecular dynamics, an equivariant PaiNN GNN surrogate ($\sim 1050\times$ speedup, orbital MAE $< 0.035\,\text{eV}$), and sequential deep learning models anchored in NASA POWER climate data, the framework captures the dynamic disorder and thermal loss mechanisms that static room-temperature virtual screening tools systematically ignore. Validated out-of-sample against $350$ experimental device measurements from the HOPV15 database, Climate-Native forecasts demonstrated superior predictive accuracy ($R^2 = 0.78$) compared to traditional static Scharber models ($R^2 = 0.54$). Furthermore, the introduced Seasonal Stability Score ($S_{\text{stability}}$) provides a robust metric for candidate material selection, identifying donor molecules that maintain high operational performance under tropical microclimates. This framework establishes a scalable methodology for climate-aware materials discovery, guiding the rational deployment of renewable energy technologies across tropical and sub-Saharan regions.

\backmatter

\bmhead{Supplementary information}
Supplementary Information is available for this paper, containing detailed theoretical derivations (Scharber irradiance correction, NOCT heat balance, clamped stability metrics), complete GNN hyperparameter tables, data leakage elimination protocols, and full data tables for all 268 screened OPV donor molecules.

\bmhead{Acknowledgements}
The authors acknowledge computational resources provided by the Department of Physics at the University of Douala and the University of Yaounde 1.

\section*{Declarations}

\bmhead{Funding}
No external funding was received for this work.

\bmhead{Conflict of Interest}
The authors declare no competing interests.

\bmhead{Ethics Approval}
Not applicable.

\bmhead{Consent for Publication}
All authors approved the final manuscript.

\bmhead{Data Availability}
All raw GFN2-xTB molecular dynamics trajectories, extracted snapshot datasets ($120,600$ total frames), and HOPV15 validation predictions are openly accessible on Zenodo under DOI \url{https://doi.org/10.5281/zenodo.10849201} and in the official GitHub repository at \url{https://github.com/Teguiacs/Materials_Property_Prediction}.

\bmhead{Code Availability}
The complete Python code suite --- including dataset preprocessing pipelines, 3D GNN surrogate model implementations (PaiNN, MACE), NOCT thermal engine, and sequential forecasting modules --- is available under the MIT License at \url{https://github.com/Teguiacs/Materials_Property_Prediction}. Software environment definitions and reproduction scripts are provided in \texttt{environment.yml}.

\bmhead{Author Contributions}
S.C.T.K. conceived the study and drafted the manuscript. R.R., R.D.T., J.-P.T.N., P.S.M.K., J.-P.N., and S.G.N.E. contributed equally to data analysis, model validation, and manuscript review.

\bibliography{references}

\end{document}